\documentstyle[12pt,epsf]{article}
\title{$K^+$ production in
$p-C$-collisions at a beam energy 1.2 GeV}
\author{A. Sibirtsev\thanks{Permanent address:
                 Laboratory of Nuclear Problems, Joint Institute of
                 Nuclear Research, Dubna, 141980, Russia
        and      Institute of Theoretical and Experimental Physics,
                 Moscow, Russia.},
K. Tsushima\thanks{Address since October:
Department of Physics and Mathematical Physics, The University
of Adelaide, Adelaide 5005 Australia.}
and Amand Faessler\\
Institut f\"ur Theoretische Physik, Universit\"at T\"ubingen,\\
Auf der Morgenstelle 14, D-72076 T\"ubingen, Germany}
\begin{document}
\maketitle
\begin{abstract}
The isobar model and the resonance model are
applied for the first analysis of
the subthreshold $K^+$-meson production
in proton-carbon collisions, which was perfomed at GSI
at an emission angle of 40 degrees and a bombarding energy of 1.2 GeV.
In this study, we focus on the role of the secondary processes
$\pi N \rightarrow K^+ Y$ in normal nuclear matter density.
It turns out that the present approach can reproduce very well
both the $\pi^+$- and $K^+$- meson spectra.
It is also demonstrated that the different kinds of descriptions for the
$\pi N \rightarrow K^+ \Lambda$ reactions substantially
differentiate the calculated results for the $p A \rightarrow K^+ X$
differential cross sections.
\end{abstract}

It has been proposed for more than one decade
that $K^+$-mesons are one of the most promising
probes for hot and dense
nuclear matter formed in heavy ion collisions~\cite{nag}.
However, in refs.~\cite{Kapusta,Cassing,Cleymans,Sibirtsev}
a quite different scenario for kaon production
in heavy ion collisions was
suggested that kaons may be produced mainly through secondary
processes in the pure hadronic phase
with nuclear matter density ($0.17 fm^{-3}$) obtained in
nucleus-nucleus collisions,
which is far away from the highly compressed and hot phase.

In order to clarify these arguments,
the $KaoS$ collaboration of GSI
have measured recently the spectra of
$\pi^+$- and $K^+$- mesons in proton-nucleus collisions
at an angle of 40 degrees and for a
beam energy of 1.2 GeV~\cite{Senger}.
It was found that the $K^+$-differential cross sections
are substantially underestimated
according to the calculations based on
only the first step processes
$p N \rightarrow K^+ \Lambda N$, even together with an inclusion of the high
momentum component of the nuclear spectral
function~\cite{deb}. Thus, it has been claimed
that the kaons might be
produced by either the secondary $\Delta-N$-
or the multi-nucleon collisions.

One of the purposes of the present study is
to analyse those data theoretically for the first time,
with particulary focusing
on the role of the secondary process $\pi N \rightarrow K Y$
in the normal hadronic phase.
We will compare our results with the GSI data for
the spectra of $K^+$-mesons
as well as those for $\pi^+$-mesons,
which will turn out to show the validity of the present approach.

For the calculations of the $\pi^+$-meson spectra
at a beam energy of 1.2 GeV, we adopt the isobar
model with the assumptions that
the $s$-wave $\Delta$-production is dominant for the
$pN \rightarrow \Delta N$ reaction, and that the $\Delta$'s decay
isotropically in the $\Delta$-isobar rest frame.
at the $\Delta$-isobar rest system.
As was demonstrated by Cugnon
et al.~\cite{Cugnon1},
the isobar model can reproduce
the experimental data on
pion production reasonably well
for beam energies below 2 GeV/u,
both in proton-nucleus and nucleus-nucleus collisions.
We use the $\Delta$-production cross section as,
\begin{equation}
\sigma (pN \rightarrow \Delta N) = \frac {Z} {A}
\sigma_{in} (pp)+ \frac {A-Z} {A} \sigma_{in} (pn),
\end{equation}
where $A$ and $Z$ denote the target mass and charge, respectively.
The  proton-proton and
proton-neutron inelastic cross sections
$\sigma_{in} (pp)$ and $\sigma_{in} (pn)$
are taken from ref.~\cite{Cugnon2}.

In order to take into account the internal momentum of the target,
we average over the $\Delta$-differential
cross section by convoluting the spectral
function $\Phi ({\bf q}$) of the target nucleus.
Then, the differential pion production cross section
in proton-nucleus collisions
can be calculated as
\begin{equation}
\label{PION}
E_{\pi} \frac {d^3 \sigma} {d^3 p_{\pi}}
(pA \rightarrow \pi X) =
N_{eff}(p_{\pi}) \ \int
\Phi ({\bf q}) \
E_{\Delta} \frac{d^3 \sigma (\sqrt{s},{\bf q})}{d^3 p_{\Delta}}
Lo_{\Delta \rightarrow  \pi N}
(\sqrt{s}, p_{\pi}) \ d {\bf q},
\end{equation}
where $E_{\Delta} d^3
\sigma (\sqrt{s}, {\bf q}) /d^3 p_{\Delta}$ stands for the
$\Delta$-differential cross section in the
center-of-mass system of the
incident proton and the target nucleus carrying the momentum
${\bf q}$ and the corresponding invariant mass $\sqrt{s}$.

In eq. (\ref{PION})
$Lo_{\Delta \rightarrow \pi N}(\sqrt{s},p_\pi)$
stands for the Lorentz transformation of the pion momentum $p_\pi$
from the $\Delta$-rest system  to the laboratory  system.

Here, the spectral function
of the target carbon
$\Phi ({\bf q})$ necessary for the present calculations
is parametrized by
the experimental data $(e,e'C)$ and $(\gamma,\gamma'C)$
scatterings as follows~\cite{Frullani};
\begin{equation}
\label{SF}
\Phi (q) = \frac {1} {\alpha^3} exp \left( - \frac {q^2}
{2\alpha^2} \right),
\end{equation}
where $q = |\vec{q}|$ and
the slope parameter $\alpha =$82 $MeV/c$. It is assumed that the
distribution of the vector ${\bf q}$ is isotropic.

The factor $N_{eff}(p_\pi)$ appearing in eq. (\ref{PION})
is accounting for the mass number $A$ dependence of
the $\Delta$ production and the final stage
of the pion absorption.
It should be mentioned that
the energy dependence of the effective collision number
$N_{eff}(p_\pi)$
must be taken into account correctly, because there is
a strong dependence of the $\pi N$ cross sections $\sigma(\pi N)$
on the produced pion momentum $p_\pi$~\cite{Cugnon2},
which can be understood clearly from Fig. 1a). Based on the approach
used in refs.~\cite{Margolis,Vercellin}, this
$N_{eff}(p_\pi)$ can be calculated as
\begin{eqnarray}
\label{eff1}
N_{eff}(p_{\pi}) & = &
\int_{0}^{+\infty} bdb \int_{-\infty}^{+\infty} \rho (b,z)dz
\int_{0}^{2\pi} d\phi \times \nonumber \\
&  & \left[ exp \left( -{\sigma}_{tot}(pN) \int_{-\infty}^{z}
\rho (b,\xi ) d \xi - {\sigma}_{tot}(\pi N, p_{\pi})
 \int_{0}^{+\infty}
\rho ({\bf r}[\zeta ]) d\zeta \right) \right],
\end{eqnarray}
where $\rho ({\bf r}[\zeta])$
is the one-particle density distribution which is taken
as a harmonic oscillator~\cite{Glauber} and
normalized to the target mass number A.
${\bf r}[\zeta ]$ appearing in eq.(\ref{eff1}) is defined as
\begin{equation}
{\bf r}[\zeta ] = {\bf r}_0(b,0,z)+ \zeta \, \hat{{\bf e}},
\end{equation}
where $b$ and $z$ stand for the impact parameter and the $z$-component of
the coordinate along the beam-axis,
respectively. $\hat{{\bf e}}$ is a unit vector in
coordinate space defined by
$(sin\theta cos \phi , sin \theta sin \phi , cos \theta )$
with $\theta$ being the detection angle. In order to obtain
the collision number $N_{eff}$ for the other processes, one may
just replace the
$\sigma_{tot}(\pi N)$ in eq.(\ref{eff1})
to the corresponding cross sections.

The effective collision number (\ref{eff1}) calculated for
the carbon target is
shown in Fig. 1b) as a function of total cross section $\sigma$.
Note that in the
calculations performed in refs.~\cite{Cassing,Sibirtsev},
the factor $N_{eff}$
was obtained by the Glauber approach, and parametrized by
\begin{equation}
\label{ALPHA}
N_{eff}=A^{0.74 \pm 0.01}.
\end{equation}
Because the total cross section for the $K^+N$-interaction is very small
$\sigma(KN) \simeq 12 mb$, one can treat it pertubatively and
can get the effective collision number $N_{eff}$
for $\sigma (KN)$
from the curve plotted in Fig. 1b).
Similar argument holds also for
$K^+$-meson production through the secondary $\pi N$-collisions,
because the reaction threshold corresponds to the pion momenta
is above 1 GeV/c and $\sigma (\pi N)$ is also small.

However, for the calculations of the $\pi^+$-meson spectra
in proton-nucelus and
nucleus-nucleus collisions one should take into acount
the pion momentum $p_\pi$ dependence of $N_{eff}$,
since the spectra reflect directly
the contributions of pion absorption.

At first we will discuss the results for $\pi^+$-meson production.
The solid line in Fig. 2a)  shows
the calculated spectrum for $\pi^+$-mesons in proton-carbon collisions
obtained by using eq. (\ref{PION}).
The experimental results are represented by the circles.
The dashed line shows the results
obtained by the phase-space calculations
for the $\pi N \rightarrow N N \pi$ reactions and averaged over
by convoluting the spectral function (eq. (\ref{SF})) and
normalized to $N_{eff}$.
Here the $\pi^+$ production cross section is given by
VerWest and Arndt~\cite{VerWest} as follows;
\begin{equation}
\sigma (pN \rightarrow NN \pi^+) =
\frac {Z} {A} \sigma (pp \rightarrow pn \pi^+) +
\frac {A-Z} {A} \sigma (pn \rightarrow nn \pi^+).
\end{equation}

One can see that
the isobar model fits the $\pi^+$-meson spectrum
quite satisfactory.
Because the high momentum tail of the pion spectrum is
reproduced pretty well and it implies that sufficient amount of
high momentum pions exist,
thus one can expect that
$K^+$-meson production through the $\pi N \rightarrow K^+ Y$
reactions is quite possible.

Next, we will discuss about $K^+$-meson poduction.
The differential kaon production cross section in proton-nucleus
collisions is given by
\begin{equation}
\label{KAON}
E_K \frac {d^3 \sigma} {d^3 p_K}
(pA \rightarrow K^+ X) = \int \frac {\xi (p_{\pi})}
 {\sigma_{tot} (pN)} \left[
E_{\pi} \frac {d^3 \sigma} {d^3 p_{\pi}}
\right] \ \Phi ({\bf q}) \
E^*_K \frac {d^3 \sigma (\sqrt{s})}
  {d^3 p^*_K} \frac {d{\bf p}_{\pi}} {E_{\pi}} d {\bf q},
\end{equation}
where the differential
pion production cross section
$E_\pi d^3 \sigma/d^3 p_\pi$
is taken the same form as that is
given by eq. (\ref{PION}), and
$E^*_K d^3 \sigma /d^3 p^*_K$ stands
for the differential kaon production cross sections
for the $\pi N \rightarrow K^+ Y$ reactions
in the center-of-mass of the
pion and the nucleon with the corresponding
invariant mass $\sqrt{s}$
of the system.
Here, we may take into account only
the $\Lambda$-hyperon production reaction channel, which is
reasonable for the small values of $\sqrt{s}$.

In eq. (\ref{KAON}) the factor $\xi(p_\pi)$
stands for the probability of the $\pi$-meson
interaction with nucleons inside the target,
and it is given by~\cite{Margolis}
\begin{equation}
\xi (p_{\pi})=1- N_{eff}(p_{\pi})/N_{eff}(\sigma(\pi N) = 0).
\end{equation}

It was found in ref. ~\cite{Cugnon3} that the angular distribution
of $K^+$-mesons induced by the $\pi N \rightarrow K^+ \Lambda$
reactions is not isotropical at all
even at energies close to the threshold.
This does not have a large
influence on the calculated results
for the total
kaon production cross sections~\cite{Cassing,Sibirtsev},
however, this anisotropic effects should be
taken into account adequately
when the calculations for the differential $K^+$-meson spectrum
are performed.

The angular spectra of $K^+$-mesons induced by the
reaction $\pi^0 p \rightarrow K^+ \Lambda$ in the center-of-mass
system are given in Fig. 3.
The results represented by the circles, squares, triangles and stars
are obtained by adopting
the resonance model ~\cite{Tsushima1,Tsushima2},
while the solid-, dashed-, dotted- and dashed-dotted- lines represent the
results obtained by the use of the
parametrization of refs.~\cite{Cugnon3,Sibirtsev1}
for the range of $\sqrt{s}$ from 1.62 to 1.7 $GeV$.
The explicite parametrization is given by,
\begin{equation}
\label{COS}
\frac {d \sigma} {d \Omega^*} \sim 1 + \eta (\sqrt{s})
cos \theta^*_K,
\end{equation}
with $\eta(\sqrt{s}) =10.9 \sqrt{s}-17.6$ and $\eta=1$, corresponding to
the ranges threshold $< \sqrt{s} <$ 1.7 $GeV$
and $\sqrt{s} >$ 1.7 $GeV$, respectively.

It is worth noting that the resonance model shows a
strong anisotropy
for $K^+$-meson production, whereas the function
represented by eq. (\ref{COS}) is quite flat
as a function of $cos \theta$ with $\theta$
being the angle between the kaon and the pion momenta.

Now we discuss the results for $K^+$-mesons.
The results are given in Fig. 2b) for
the kaon spectra in proton-carbon collisions at a kaon angle of 40 degrees
and a proton bombarding energy of 1.2 GeV
by applying eq. (\ref{KAON}).
The solid line shows the
results obtained by adopting the resonance model, while the
dashed line shows the results obtained with eq. (\ref{COS}). The
experimental data are given by the circles.
The calculations preformed by adopting the
resonance model can reproduce the experimental data quite satisfactory.
This fact indicates that the importance of the precise
description for the $\pi N \rightarrow K \Lambda$ reactions.
Furthermore, this leads
support to the assumption in the present calculations,
that $K^+$-mesons may be produced
in a pure hadronic phase under the normal conditions.

To summarize, we have performed the first theoretical analysis
for $K^+$-meson production in proton-carbon collisions which is
measured at GSI~\cite{Senger}.
In the calculations, the isobar model and the resonance model
were adopted with
normal nuclear matter conditions.
By taking into account the experimental spectral function for the
carbon target,
we could reproduce the experimental spectra
quite well for both the
$\pi^+$- and $K^+$- mesons.
It was demonstrated that the secondary $\pi N$-collision  processes
give dominant contributions for $K^+$-meson production in
proton-nucleus collisions below the reaction threshold in free space
as was discussed by Cassing et al.~\cite{Cassing2}.
The present results support the scenario that the $K^+$-mesons
may be produced mainly in the pure hadronic phase under the normal
conditions through the secondary
processes~\cite{Kapusta,Cassing,Cleymans,Sibirtsev}
$\pi N \rightarrow K Y$.
Furthermore, it turned out that the angular distribution
of the $K^+$-meson
spectrum in proton-nucleus collisions is quite sensitive to
the $\pi N \rightarrow K \Lambda$ differential cross sections,
and can serve to differentiate between the two models.
We therefore urge the experimental group at GSI to
measure such an angular distribution of the kaons. For this
purpose we give in figure 4 the angular distribution of
the kaons integrated from the kaon momentum
$p_K = 0.5$ to $0.7$ Gev/c for
the reaction $^{12}C(p,K^+)$ at 1.2 GeV proton
bombarding energy.\\

\noindent{\bf Acknowledgement}
The authors express their sincere
thanks to E. Grosse, P. Senger and M. Debowski
for allowing us to use the experimental data prior to
their publication.

\newpage
\vspace*{2cm}
\noindent{\Large {\bf Figure captions}}\\ \\
\noindent Figure 1: Total $\pi^+p$ cross section as a function of
pion momentum $p_\pi$  a), and the
calculated effective collision number
for $^{12}C$ as a function of
total cross section $\sigma$ b).\\ \\

\noindent Figure 2: The spectra of $\pi^+$- and
$K^+$- mesons in proton-carbon
collisions at an angle of 40 degrees
and a beam energy of 1.2 GeV. The circles in Figs. stand for
the experimental data.
The solid- and the dashed- lines in Fig. a)
show the results obtained by
the isobar model and the phase space distributions, respectively.
The solid- and the dashed- lines in Fig. b) represent the results
obtained by adopting the resonance model
\cite{Tsushima1} and
 eq. (\ref{COS}), respectively.\\ \\

\noindent Figure 3: The angular spectra of $K^+$-mesons
for the $\pi^0 p \rightarrow K^+ \Lambda$ reaction.
The different curves are plotted as a function of the
total center of mass energy $\sqrt{s}$.
They represent the following theoretical results:
The circles and the solid line
give the results for
$\sqrt{s}$ =1.62 GeV,
the squares and the dashed line for
$\sqrt{s}$ =1.64 GeV,
the triangles and the dotted line for
$\sqrt{s}$ =1.66 GeV,
and the stars and the dashed-dotted line for
$\sqrt{s}$ =1.70 GeV, respectively.
Here, the circles, squares, triangles and stars
show the results obtained
by applying the resonance model,
while the solid-, dashed-, dotted- and dashed-dotted-
lines show the
results obtained with eq. (\ref{COS}).\\ \\

\noindent Figure 4: The angular
distribution of kaons with the experimental data
for the $^{12}C(p,K^+)$ reaction at 1.2 GeV proton
bombarding energy. The results shown in figure are
obtained by integrating the kaon
momentum from $p_K = 0.5$ to $0.7$ GeV/c.
The solid-
and the dashed- lines show the results obtained
by adopting the resonance model~\cite{Tsushima1}
and  eq. (\ref{COS}), respectively.

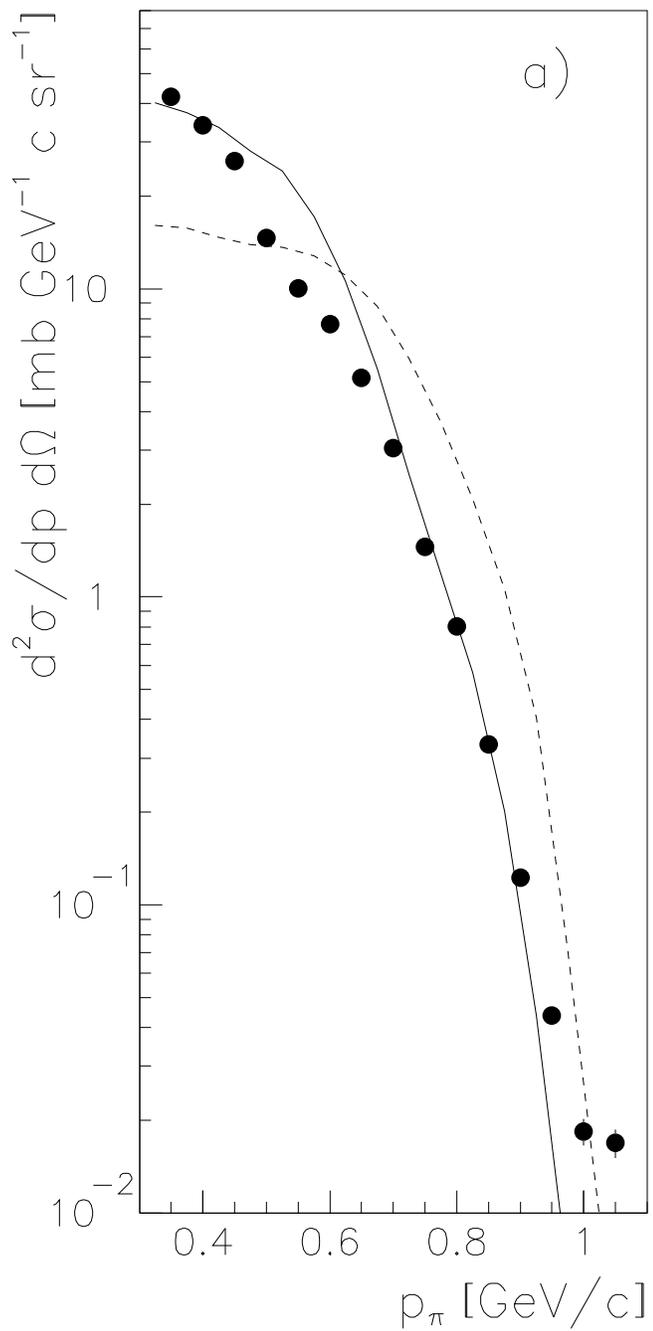
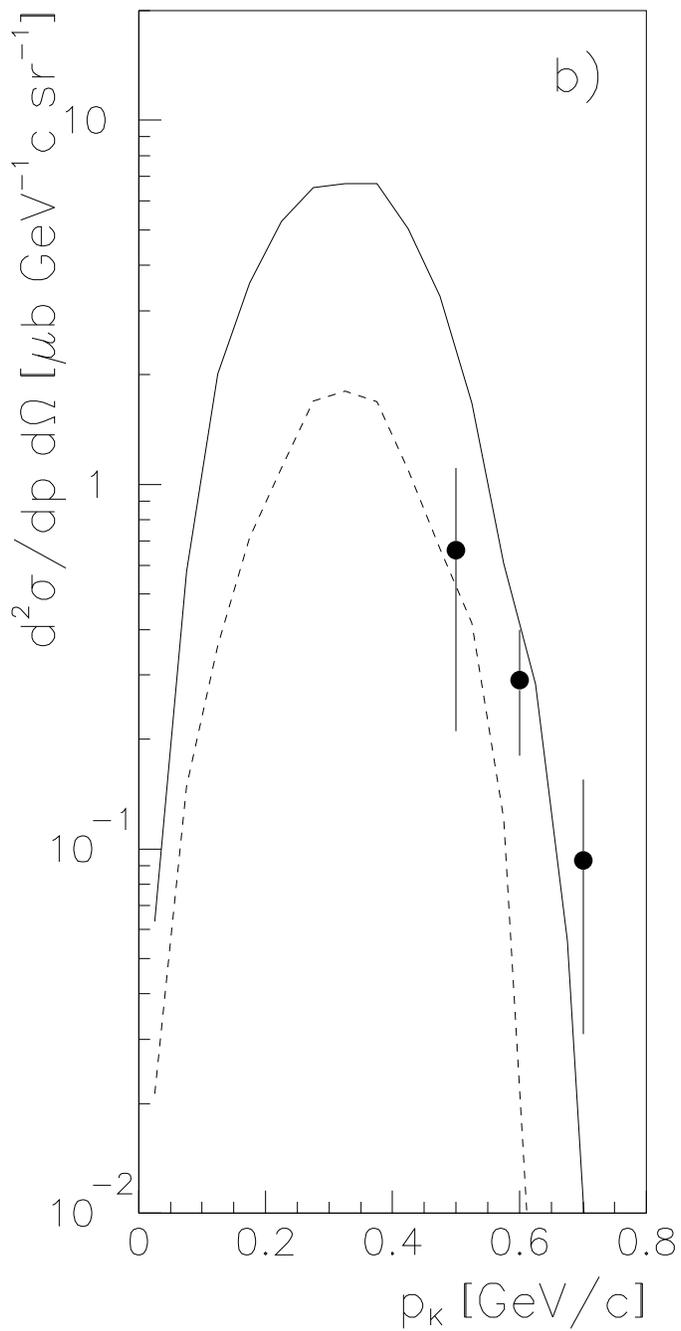
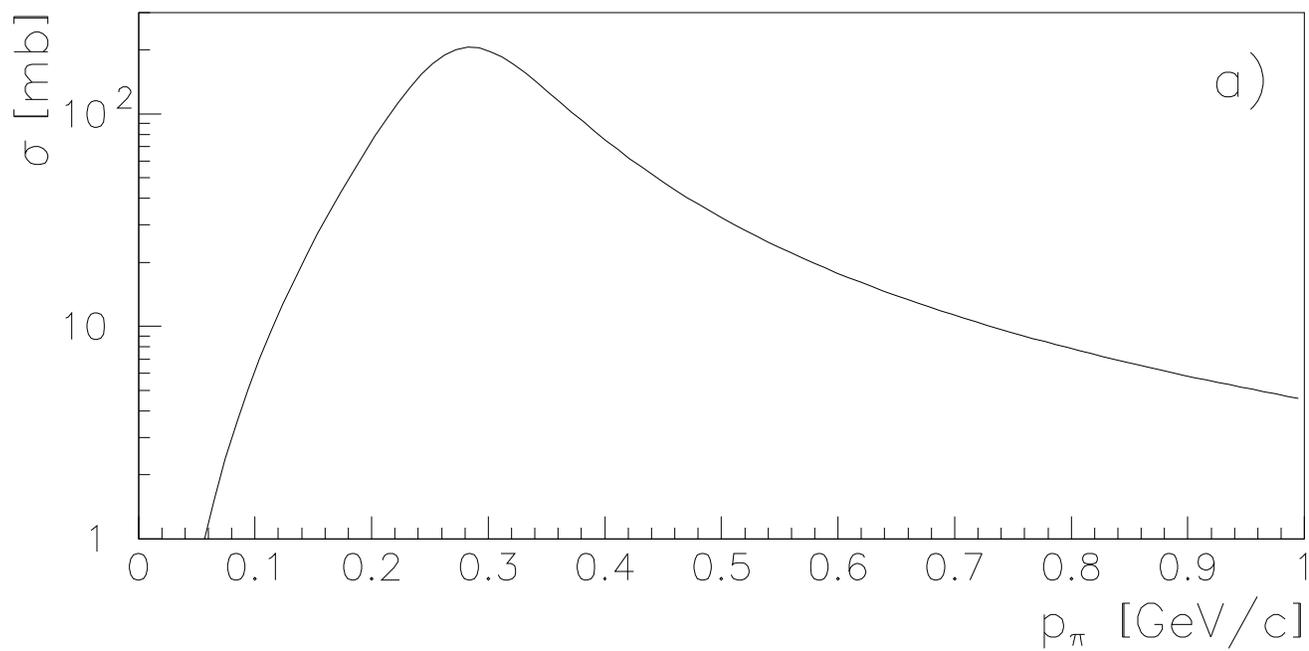
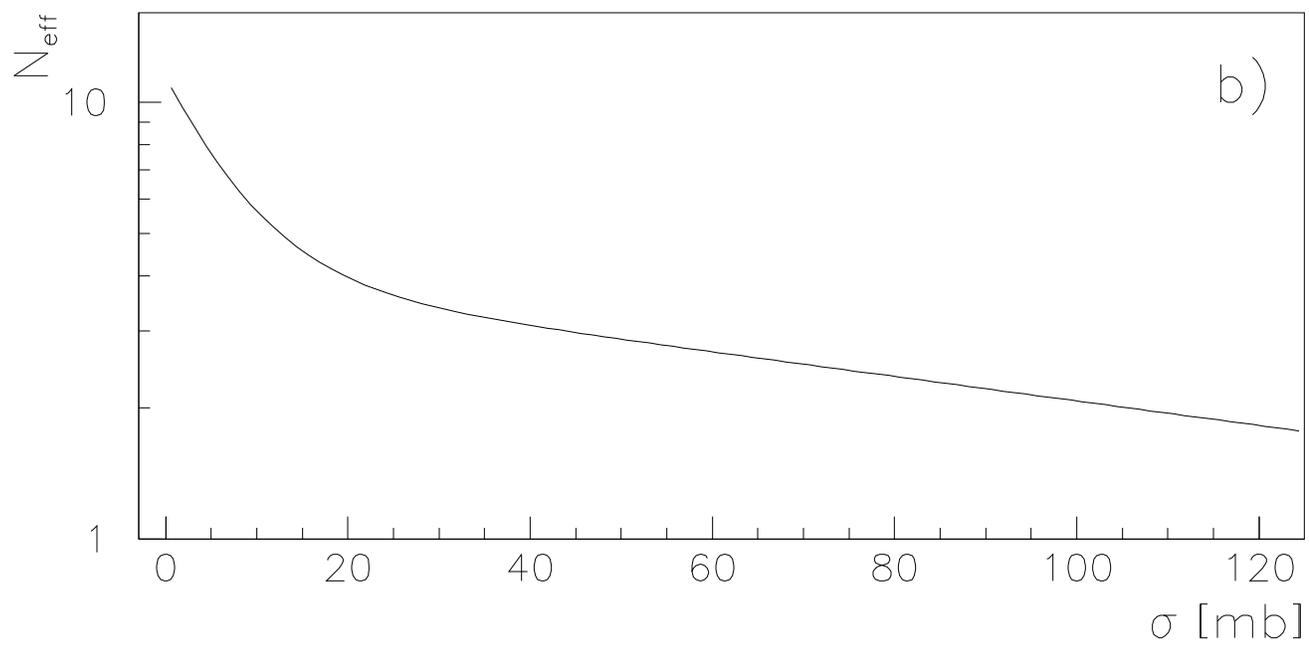
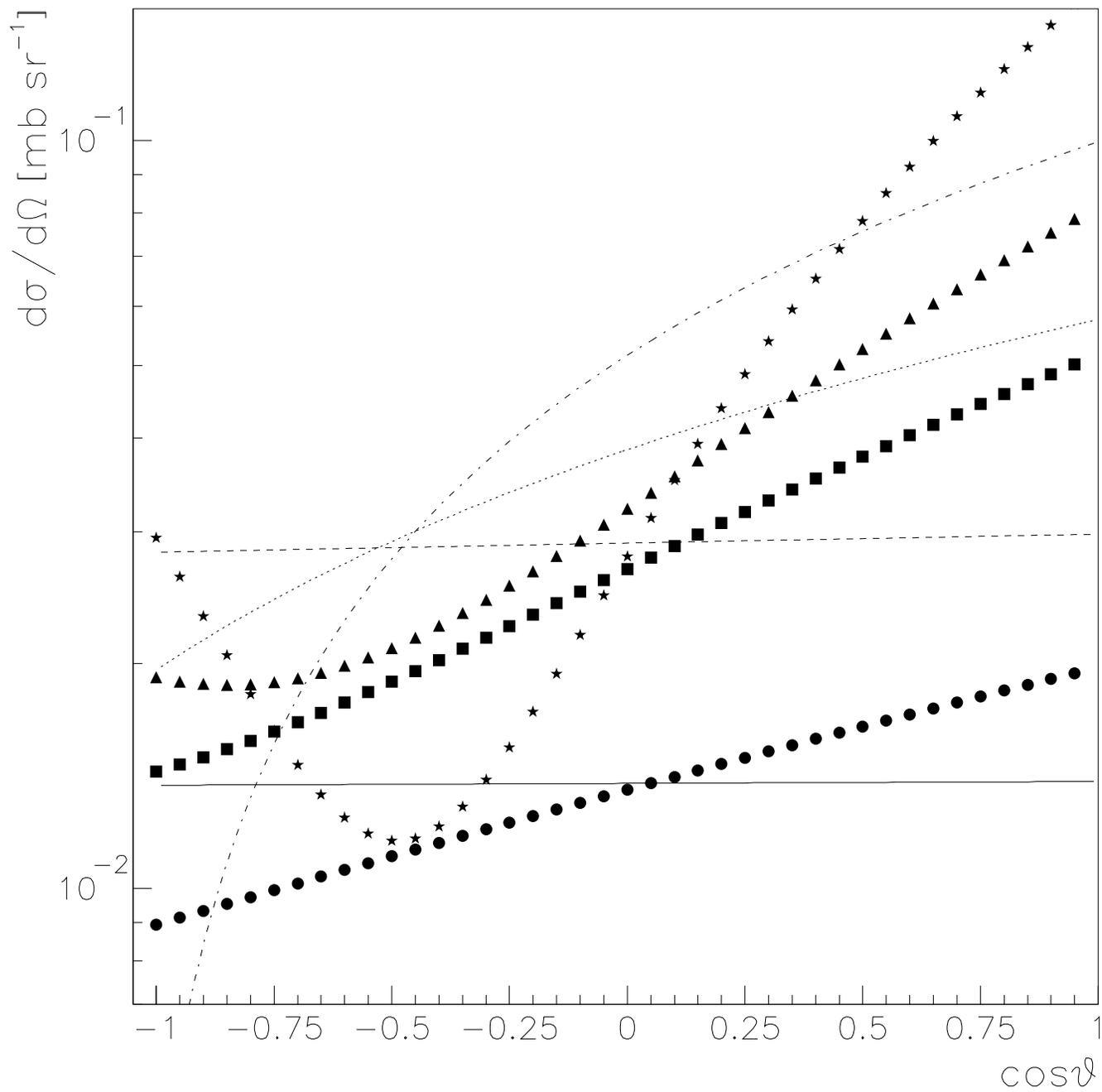
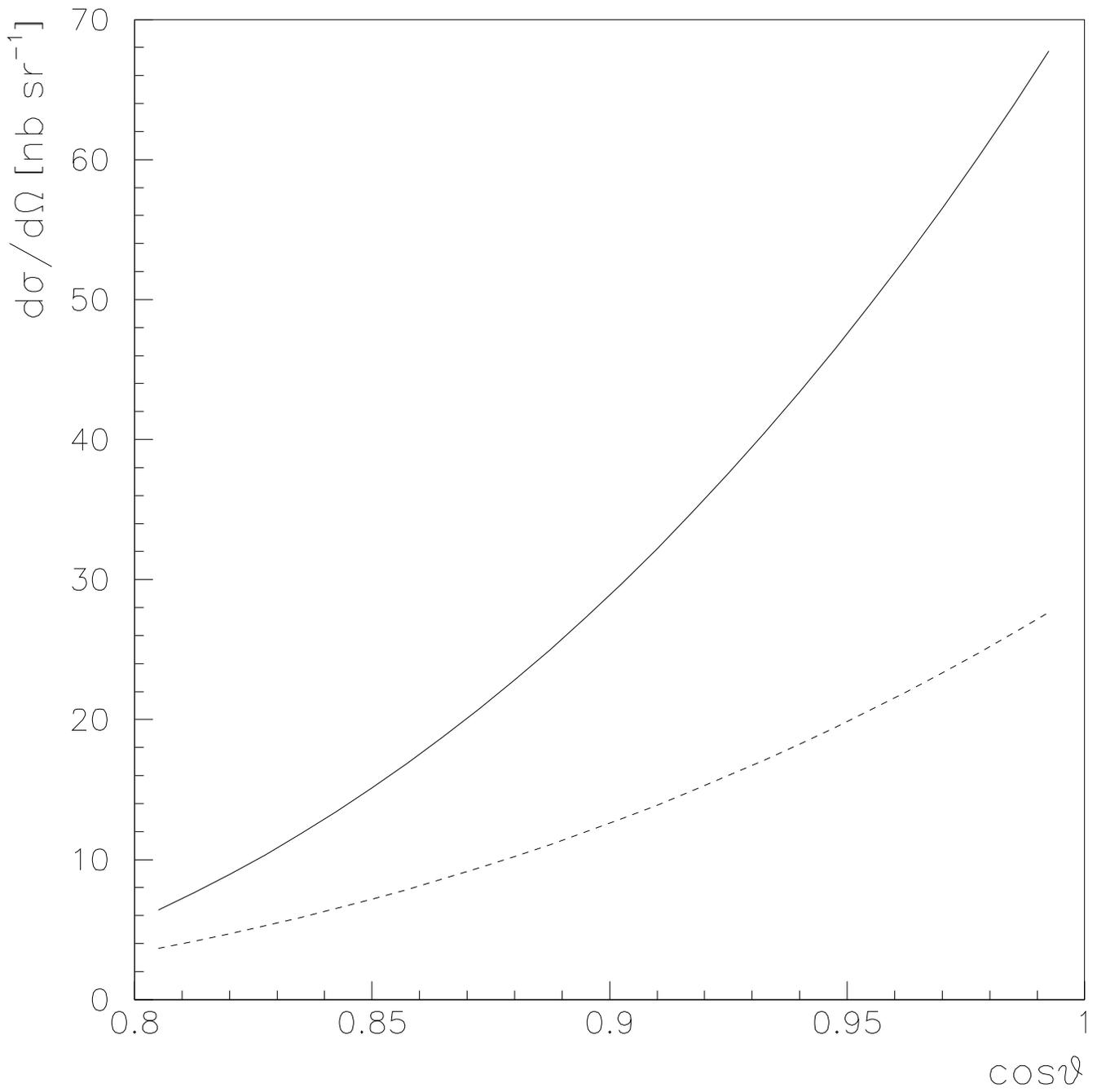
\end{document}